# Environmental policies as a pull factor for tourists? Insights from Italy

Riccardo Gianluigi Serio[1*], Maria Michela Dickson[1], Thomas de Graaff[2,3], Eric H. Pels[2,3]

[1] Department of Economics and Management, University of Trento, Italy
[2] Department of Spatial Economics, Vrije Universiteit Amsterdam, The Netherlands
[3] Tinbergen Institute, Amsterdam, The Netherlands

[*]Corresponding author: Riccardo Gianluigi Serio. Email: riccardo.serio@unitn.it

### Abstract

Tourism consumption has grown into a major economic factor for modern societies. However, the environmental impact of tourism has become a significant concern, leading to an increased focus on sustainable tourism policies. While governments and institutions have introduced frameworks to promote ecological transition in the tourism sector, the effectiveness of such policies remains unclear. This study provides a seminal attempt to examine the complex relationship between tourism demand and sustainable tourism policies. To do so, a gravity model framework has been adopted to examine incoming international tourism flows in Italian provinces in 2019. The findings reveal a positive association between tourism demand and sustainable labels. This study also suggests that eco-labels are appreciated by tourists and have a role in the destination decision-making process. It highlighted the need for continued research to identify effective sustainable tourism policies that can balance the economic benefits of tourism with environmental considerations.

## 1 Introduction

During the last decades, tourism consumption has grown into a major economic factor for modern societies. Nowadays tourist activities constitute an important part in peoples' lifestyles due to an increasing rise in disposable free time and income (Cracolici and Nijkamp, 2008). Moreover, the tourism sector, together with tourist activities, is an important engine of economic growth (Lee and Brahmasrene, 2013), contributing to world GDP by 10.4% in 2019 (WTTC, 2021). This is particularly true for large tourist destinations such as Italy, in which tourism contributes attributes 7% to GDP, and is responsible 7.1% of jobs (Borsaitaliana, 2021).

The preferences of tourists have changed over the years, and new and more detailed elements come into play as determinants of the final choice of the destination. Several studies have been carried out to identify the reasons behind the success of a tourist destination (see, among others, Alavi and Yasin, 2000; Crouch and Ritchie, 1999; Enright and Newton, 2004; Kozak and Rimmington, 1999; Kozak, 2002; Ruhanen, 2007), moving from a vision in which the main elements were distinctly constituted by the destination's environmental, cultural, and historical aspects, to a more homogeneous concept, in which the destination is conceived as an evolved and multi-offering supply of the most various products and services (Cracolici and Nijkamp, 2008; Buhalis, 2000; Ritchie and Crouch, 2000). The actors in this network have the task of creating a coherent and adequate offer combining the resources that the destination provides, to gain a competitive advantage over the experiences offered by other competing destinations (Teece et al., 1997).

Strongly entwined with tourism development is environmental sustainability, a topic that has become increasingly more important, as well in relation to tourism. Concepts such as sustainability landed in tourism policy and decision making at the end of 1980s, when the idea of sustainable development was introduced to tourism (Hall, 2011). Both governments and national and supra-national institutions have worked hard to provide frameworks that allow for an ecological transition in the tourism sector (e.g., the European Union with the European Charter for Sustainable Tourism in Protected Areas, 1995; ACI Europe with the Airport Carbon Accreditation Program, 2009; ICAO, 2011; United Nations Environmental Program, 2003; The European Tourism Indicator System: ETIS toolkit for sustainable destination management, 2017). However, the effectiveness of such policies remains unclear, and some scholars have pointed out the need to strengthen the research on sustainable tourism policies (Guo et al., 2019).

At the same time, researchers have enriched the literature on environmental sustainability policies related to tourism in several respects. For instance, in-depth literature reviews on the environmental sustainability of tourism policies, point out that although the attention of academics on the topic has grown considerably in recent years, there are still remaining challenges in identifying and applying sustainable policies, starting with the ambiguity surrounding the mere definition of sustainable tourism policies (see, e.g., Guo et al., 2019; Torres-Delgado and Palomeque, 2012). Others have created a theoretical framework of applicability for sustainable tourism policies, distinguishing between policies geared towards reinvigorating the image of the destination that contain elements of sustainability (misaligned from the true goal of the ecological transition of tourism) and policies focused on the preservation of the natural environment (more effective in reducing the environmental impact of the sector) (Farsari et al., 2007). Finally, scholars focused their research on proposing complex indicators for the assessment of environmental impacts of tourism related activities, in order to progressively mitigate the negative environmental externalities due to tourism activities (see, e.g. Castellani and Sala, 2010; Blancas et al., 2010).

This paper aims to link several tourisms environmentally sustainable initiatives (such as green labels) to tourism flows. To do so, we adopt a gravity model framework to investigate incoming international tourism flows in Italy in 2019 at a provincial level of dis-aggregation. This framework allows us to implement a first attempt in understanding if environmentally sustainable initiatives in tourism act as an inhibitor or facilitator for tourism flows. Our findings reveal a clear positive association between tourism demand and sustainable labels and suggest that the importance of private and firms' commitment is larger than public efforts.

The paper is structured as follows. The next section presents a theoretical framework to link gravity specification and tourism sustainable policies studies. We present the data in Section 3, whereafter Section 4 discusses the results. The last section concludes.

## 2 Theoretical framework

In recent decades, environmental sustainability and sustainable transition have emerged as topics of interest in several research streams, including tourism. Tourism can significantly and consistently contribute to the economic growth of the hosting region (WTTC, 2021), but, at the same time, it also has significant impact on the environment (Weaver, 2012; Weaver, 2014). To deal with environmental challenges, national governments and supranational institutions, such as the European Union and the United Nations, have developed several measures focused on reducing environmental pollution driven by tourism activities. Some notable examples are the European Charter for Sustainable Development (2003) and the ETIS toolkit (2019). However, the extent to which such policies, programs, and labels effectively reduce tourism pollution remains unclear and partially unexplored. In the following sections, we first give an overview about some initiatives adopted in recent years by national and transnational governmental bodies to push tourism toward the green transition. Thereafter, we present a literature review about the efforts in understanding the effectiveness of such practices.

### 2.1 Sustainable tourism policies

One of the first attempts in implementing practices for sustainable tourism was made in 1992, when the European Union (EU) introduced the ecological labeling system (on a voluntary basis) Ecolabel EU with regulation 800/92, The ecolabel certifies the reduced environmental impact of products and services considering all phases of the life cycle, through selective criteria defined on a scientific basis. Additionally, the EU introduced EMAS (Eco-Management and Audit Scheme), which allows public and private companies with headquarters based in an European country to improve their environmental performance.

Companies can also voluntary adopt certification measures, which ensure quality standards much higher than those set by law such as ISO 9001, ISO 14001. For coastal areas, the Blue Flag label (BF), whose standards are enshrined in the Foundation for Environmental Education (FEE), has been designed to improve sustainability in coastal destinations. Regarding the airport sector, the Airport Carbon Accreditation (ACA) is the only globally institutionally recognized carbon footprint reduction label. The program consists of six certification steps: Mapping, Reduction, Optimization,

Neutrality, Transformation, and Transition. Launched in 2008 by ACI (Airport Council International), ACA accredits airports' commitment to undertake a virtuous path that culminates in achieving emission neutrality, also considering energy management. Compliance with ISO14064 (Greenhouse Gas Accounting) is a fundamental prerequisite for getting access to this program. With the attainment of the last level of accreditation, namely "Transition", the airport proves to be aligned with the 2015 Paris agreements, actively contributing to curbing the increase of the global average temperature to 1.5° C and no more than 2° C compared to the pre-industrial levels.

In addition to public policies such as EMAS and Ecolabel, private companies have also created systems aimed at reducing environmental impacts, particularly in the tourism sector. BioHotels, for example, is a label developed in the early 2000s by a group of hoteliers with the aim of providing high-quality tourist services with zero environmental impact. Today, there are over 80 hotels that are part of BioHotels, 13 of which are in Italy. Another label synonymous with sustainable tourism is Green Key, which has over 3,000 registered structures and is a prestigious label issued by FEE (Foundation for Environmental Education), a non-governmental and non-profit organization based in Denmark. Among FEE's strategic partners, UNEP (UN Environment Program) and UNWTO (UN World Tourism Organization) are the most notable. Moreover, the Green Key program aims to reduce water resource consumption, manage waste, and use electricity from renewable sources to reduce overall environmental impact and costs.

Although the adoption of such standards represents a step towards greater awareness of environmental problems, it is not clear what the effect of such initiatives and labels is on tourism flows and, consequently, on the economic growth of the hosting region. The literature discussed in the next subsection suggests that, in general, environmental tourism policy may not be that effective.

## 2.2 Literature review

Several researchers suggest that the lack of a clear definition of "sustainable tourism policy" is one of the main reasons behind the failure to implement effective environmental policies in the tourism industry (e.g., Guo et al., 2019). Sustainable tourism policies often appear vague and malleable (Farsari et al., 2007), with a mainly economic focus (Yüksel et al., 2012), making them difficult or impossible to apply in practice and to be evaluated ex-post. Authors have also noted that the evaluation of externalities resulting from new policies is severely limited by the scarcity and fragmentation of collected data during follow-up (Farsari et al., 2007) and by the complexity of the guidelines set by policymakers. This complexity is partly linked to the nature of tourism policies, which inevitably influence and are influenced by other policies (Dodds and Butler, 2009). Although governments play a crucial role in fostering the sustainability of a destination (Guo et al., 2019), their leadership is sometimes unable to guide tourism toward ecological transition (Andersen et al., 2018). Another problem discussed in the literature is the divergence of time horizons. Despite governments' declaration of commitment to guide the ecological transition, which is structurally a long-term goal, the proposed policies usually have short time horizons because short-term economic interests outweigh long-term environmental interests (Dodds, 2007). Moreover, several researchers have studied the role of sustainable tourism in promoting regional development through the construction of sustainability indicators (Miller, 2001; Miller and Ward, 2005; Castellani and Sala, 2010; Garcia and Staples, 2000), while others focus the attention on understanding if an increase in attractiveness of specific areas can be due to standards and labels (Bernini and Cerqua, 2020; Cerqua, 2017; Zielinski and Botero, 2019). The current paper is framed in this stream of research, and aims to investigate the relationship between environmentally sustainable initiatives and the size of tourism flows.

There is already a large literature on the size of tourism flows and their determinants, and it can be divided in two streams. On the one hand, tourism determinants are studied focusing on changes over time, employing time-series methods (e.g., Crouch, 1994; Song et al., 2019). On the other hand, if the interest is switched from time to space, such as in cases where there are many destinations or many origin countries, gravity models are usually employed. Gravity models have been widely used in international trade literature since the seminal work of Tinbergen in 1962. However, their application to tourism flows has been relatively recent. While Zipf (1946) studied tourism flows between two cities in the U.S., it was not until the 1970s that the first paper specifically addressing tourism movements was published (Armstrong, 1972). The interest for gravity models in the tourism sector suffered ups and downs over time, due to the complexity of tourist behaviour that made these models difficult to apply (see e.g., Colwell, 1982; Morley et al., 2014). However, gravity models are still being used as a solution for tourism demand modeling, using GDP, population, distance and many other variables, including sustainability, to explain tourism flows. Despite some recent attempts in literature (see e.g., among others, Massidda and Etzo, 2012; Xu and Dong, 2020) that include environmental variables in the model, no previous studies have tried to use composite environmental indicators as independent variables. In the present study we proceed in this direction in order to verify if local sustainable practices relate to more sustainable tourism consumption.

# 3 Gravity models for tourism flows

Morley et al. (2014) offers an approach that overcome the lack of theoretical foundation in tourism studies and improve its application to policy decision-making. They propose a gravity model assuming that the destination qualities influence the individual's utility function, or, the utility of an individual in visiting destination $j$ moving from origin $i$ in year $t$ ($U_{i,j,t}$) *i*s a function of: (*i*) the total number of visits that specific person visits $j$ moving from $i$ in a timespan $t$ ($NV$), (*ii*) other products and services consumption in the origin ($OP$), (*iii*) origin-related characteristics of dimension $s$ ($O^s$), and (*iv*) destination-related characteristics of dimension $p$ ($D^p$).

Such a utility specification can then, in general, be expressed by:

$$U_{i,j,t} = f(NV_{i,j,t}, OP_{i,t}, O^s_{i,t}, D^p_{j,t}) \qquad (1)$$

Utility is maximized subject to the constraint that the individual will visit a destination until the cost of visiting $j$ from $i$ in $t$ summed to other goods costs do not exceed the individual's available budget. As such, Morley et al. (2014) demonstrate that the aggregated demand function can be derived from the utility function in Equation (1).

The main theoretical problem that Morley et al. (2014) highlight, lies in the fact that previous research, due to the perishable nature of tourism products, mainly focused on forecasting and time series-based approaches. Thus, most applications of gravity models in tourism demand contexts have been restricted to forecasting problems based on income and prices. In short, given that the origin and destination determinants (hereafter, $O$ and $D$), are not subject to significant changes in small time horizons (i.e., one year), these two components of the equation lose their importance and interest in favor of other determinants (which instead change in the short run), such as price elasticities and income (see e.g., Crouch, 1994; Song and Li, 2008). However, when the research focus shifts toward origin and destination determinants ($O$ and $D$), the entire Equation (1) can be validly studied via its spatial dimension ($i, j$).

We can rewrite Equation (1) into Equation (2), which represents the classical functional form for modeling the aggregate demand for tourism. That is, the power model of Witt and Witt (1995):

$$NV_{IJ} = \prod_{s=1}^{S}(O_I^s)^{\alpha_s} \cdot \prod_{p=1}^{P}(D_J^p)^{\beta_p} \cdot \prod_{r=1}^{R}(OD_{IJ}^r)^{\xi_r} \qquad (2)$$

Taking natural logarithms yields Equation (3), after enabling again the temporal dimension.

$$lnNV_{IJt} = \alpha_0 + \sum_{s=1}^{S}\alpha_s \cdot lnO_{It}^s + \sum_{p=1}^{P}\beta_p \cdot lnD_{Jt}^p + \sum_{r=1}^{R}\xi_r \cdot lnOD_{IJt}^r , \qquad (3)$$

where the dependent variable $NV$ denotes the flow of tourists between two or more region, or, more simply, the aggregate tourism flow. On the right-hand side, the explanatory variables are grouped into three main categories: in $O$, the pushing forces determining the outbound tourism in the origin are collected (e.g., GDP, accessibility to (air-)transport, population). Then, in $D$, the pulling forces able to attract visitors to a given destination, are grouped (e.g., GDP, population, museums and cultural attractions, natural attractions). In this paper, we also consider the provision of sustainable infrastructure and the adoption of policies aimed at reducing the environmental impact of tourism activities as determinants of tourist utility. Finally, in $OD$, origin-destination specific characteristics are presented (such as distance).

Since the aim of this paper is to investigate the relationship between tourist flows and environmental sustainability we build a model to control for many factors that can potentially confound the effects of our sustainability variables on tourism flows. Moreover, we give more attention to pull factors, since we investigate tourism environmental efforts of the destination. Therefore, we apply a spatial fixed effects specification to reduce the disturbance arising from omitted variable issues and unobserved heterogeneity, following Kuminoff et al. (2010), who suggest that fixed effects are to be considered the preferable approach to reduce spatial heterogeneity in cross-sectional data.

Regarding the computational method, several contributions have been proposed in literature, spanning from Poisson pseudo-maximum likelihood Silva and Tenreyro (2006) to Ordinary Least Square (OLS) estimators. Despite all the benefits of the first, OLS remains the most widely used method for estimating such models Rosselló Nadal and Santana Gallego (2022). In this paper, we have therefore adopted an OLS-estimated gravity model (in its log-log form), in which we have included country of origin fixed effects to partially control for endogeneity and use heteroskedasticity consistent standard errors.

We compute four models of increasing complexity, starting from a basic gravity specifi- cation with two masses (GDPs) and distance. The computational method is OLS with White heteroskedasticity-consistent standard errors. Also, the distance and most other regressors (except for the a dummy for "Domestic flows" and the 2 count measures of sustainable initiatives) are taken in the logarithmic form for sake of interpretability, to ensure consistency with the theoretical underpinning under gravity specifications in tourism literature (i.e., Morley et al., 2014), and to avoid specification bias in the relation between tourism flows and all the independent variables

# 4 Data description and presentation

Gravity models on tourism demand traditionally use as dependent variable tourist arrivals, presences, trips, bilateral money flows and receipts (i.e. Petit and Seetaram, 2019; Mata and Llano-Verduras, 2012). In the present paper, the total number of overnight stays (that is, total presences) in all tourism accommodations, for 107 provinces in Italy in 2019 is used. This choice is due to data availability at a NUTS 3 level, for which tourism expenditure flows are not publicly available. The dependent variable has been built using data coming from the Italian National Institute of Statistics (ISTAT), which releases tourist flows from world countries (NUTS 1) to Italian provinces (NUTS 3). Because the purpose of the current paper is to understand the role of environmental initiatives in promoting tourism flows, we consider only European countries as the origin, so that origins and destinations fall under a comparable legislative and social-economic framework.

Distance is usually considered as an independent variable, both in geographical and travel terms. The expected effect of distance on tourism flows, as previously mentioned, is negative, implying that the greater the distance between two countries, the lower the expected utility, and thus the lower the tourism flow. For our purpose, we compute the distance between two points in a Cartesian space. We consider as well the GDP per capita of both destination and origin, in order to capture the probability to travel or the propensity for trips due to greater average income levels.

To control for population, we incorporate a population density measure for the destination, which we expect to be negatively correlated with tourism flows, as shown by Marrocu and Paci (2013) for domestic tourism in Italy. We also incorporate the unemployment level of the destination to examine production levels and a variable that accounts for the educational level of the destination (namely the total number of graduated students per province for each year).

Another category of explanatory variables frequently contemplated in the literature is geographical and accessibility-related variables, such as levels of infrastructure, air connectivity, and common border sharing (Rosselló Nadal and Santana Gallego, 2022). Note that 69 of the 107 Italian provinces do not offer aviation services. We include as well a variable that explains the accessibility to each province. These data sources were retrieved from Espon databases and account for all transport modes' potential accessibility. The expected relationship with accessibility variables is positive. The higher the potential accessibility, the easier to access transportation services, with consequent higher probability that people will travel, higher accessibility in the destination, and the easier it is for a tourist to reach the province and access to neighboring areas.

Further, environmental and climate variables have been also included in previous research (Rosselló Nadal and Santana Gallego, 2022). Marrocu and Paci (2013) show that Blue Flag labels for coastal destinations exhibit a positive association with higher tourism flows. Massidda and Etzo (2012) investigated the effects of environmental variables on domestic bilateral flows between Italian Regions (at a NUTS-2 level), finding that environmental facets are more important for tourism in southern destinations, while northern ones are preferred for cultural-related activities. Finally, Xu and Dong (2020), in studying the effects of air pollution effects on inbound tourism flows in China, found that both the origin and destination regional air pollution negatively impacts tourists' arrivals. Hence, we included several environmental-related variables grouping the information into two comprehensive variables, aiming at investigating the effect of environmental sustainability policies on tourism inbound flows in Italian provinces.

The first variable, *"Environmental initiatives-provinces"*, captures several green labels and sustainable initiatives aimed at improving environmental performances at the municipal and public authority level. The information considered are the total number of Blue Flags, and the total number of public administrations and waste management firms labeled with the EMAS label located within the province. The second variable, namely *"Environmental initiatives-firms"*, accounts instead for environmental performances of firms operating in the tourism industry inside the province, such as airports, hotels and cleaning services. Here the lables considered in building the variable are EMAS accommodation, Ecolabel accommodation, Ecolabel other firms and products, Ecolabel cleaning services, Airport Carbon Accreditation, Bio Hotels and Green Key projects.

The analysis is conducted on 3,424 origin-destination observations from 32 European States to each one of the 107 Italian provinces. In Table 1, the list of the 31 origin countries with geographical coordinates and GDP is reported. Switzerland and Liechtenstein are grouped together as a single entity because tourism flows are aggregated. Also, GDP for these origin countries is averaged.

|    | Country                      | Lat   | Long   | $GDP_{per\ capita}$ |
|----|------------------------------|-------|--------|---------------------|
| 1  | Austria                      | 47.33 | 13.33  | 44740.00            |
| 2  | Belgium                      | 50.83 | 4.00   | 41660.00            |
| 3  | Bulgaria                     | 43.00 | 25.00  | 8820.00             |
| 4  | Cyprus                       | 35.00 | 33.00  | 26280.00            |
| 5  | Croatia                      | 45.17 | 15.50  | 13680.00            |
| 6  | Czechia                      | 49.75 | 15.50  | 21150.00            |
| 7  | Denmark                      | 56.00 | 10.00  | 53210.00            |
| 8  | Estonia                      | 59.00 | 26.00  | 19940.00            |
| 9  | Finland                      | 64.00 | 26.00  | 43440.00            |
| 10 | France                       | 46.00 | 2.00   | 35970.00            |
| 11 | Germany                      | 51.00 | 9.00   | 41800.00            |
| 12 | Greece                       | 39.00 | 22.00  | 17100.00            |
| 13 | Hungary                      | 47.00 | 20.00  | 15000.00            |
| 14 | Ireland                      | 53.00 | -8.00  | 72400.00            |
| 15 | Italy                        | 42.83 | 12.83  | 30080.00            |
| 16 | Latvia                       | 57.00 | 25.00  | 16040.00            |
| 17 | Lithuania                    | 56.00 | 24.00  | 17500.00            |
| 18 | Luxembourg                   | 49.75 | 6.17   | 100360.00           |
| 19 | Malta                        | 35.83 | 14.58  | 27830.00            |
| 20 | Netherlands                  | 52.50 | 5.75   | 46880.00            |
| 21 | Norway                       | 62.00 | 10.00  | 67640.00            |
| 22 | Poland                       | 52.00 | 20.00  | 13870.00            |
| 23 | Portugal                     | 39.50 | -8.00  | 20840.00            |
| 24 | Romania                      | 46.00 | 25.00  | 11560.00            |
| 25 | Russia                       | 60.00 | 100.00 | 10304.00            |
| 26 | Slovakia                     | 48.67 | 19.50  | 17320.00            |
| 27 | Slovenia                     | 46.00 | 15.00  | 23230.00            |
| 28 | Spain                        | 40.00 | -4.00  | 26440.00            |
| 29 | Sweden                       | 62.00 | 15.00  | 46390.00            |
| 30 | Switzerland and Liechtenstein| 47.08 | 8.76   | 110885.00           |
| 31 | Turkey                       | 39.00 | 35.00  | 8210.00             |
| 32 | United Kingdom               | 54.00 | -2.00  | 37830.00            |

**Table 1:** States of origin, with geographical location and GDP per capital levels.

In Table 2 the summary statistics for each of the variables included in our model are presented.

|  | Min | 1st Qu. | Median | Mean | 3rd Qu. | Max | SD |
|---|---|---|---|---|---|---|---|
| * Tourism flows | 0.00 | 6.99 | 8.54 | 8.63 | 10.18 | 16.61 | 2.41 |
| * Distance | -1.14 | 2.12 | 2.59 | 2.52 | 2.97 | 4.54 | 14.77 |
| *° $GDP_O$ | 9.01 | 9.73 | 10.18 | 10.21 | 10.69 | 11.62 | 0.67 |
| *° $GDP_D$ | 9.66 | 9.91 | 10.21 | 10.18 | 10.40 | 10.93 | 0.28 |
| * $Unemployment_D$ | 0.029 | 0.059 | 0.082 | 0.104 | 0.139 | 0.289 | 0.057 |
| * $Density_D$ | 3.60 | 4.65 | 5.16 | 5.18 | 5.64 | 7.86 | 0.81 |
| * $Accessibility_D$ | 3.60 | 4.24 | 4.49 | 4.46 | 4.69 | 5.01 | 0.32 |
| * Museum $visitors_D$ | 0.00 | 1.73 | 2.25 | 2.36 | 2.95 | 5.04 | 0.99 |
| * $Coasts_D$ (km) | 0.00 | 0.00 | 2.57 | 2.32 | 4.63 | 6.75 | 2.32 |
| Domestic flows | 0.00 | 0.00 | 0.00 | 0.03 | 0.00 | 1.00 | 0.17 |
| * No. tourism $firms_D$ | 5.16 | 6.34 | 6.86 | 6.91 | 7.40 | 10.33 | 0.86 |
| * Air $pollution_D$ | 2.56 | 2.89 | 3.06 | 3.09 | 3.26 | 3.57 | 0.22 |
| ° Environmental initiatives-$firms_D$ | 0.00 | 0.00 | 0.08 | 0.22 | 0.29 | 2.00 | 0.36 |
| ° Environmental initiatives-$province_D$ | 0.00 | 0.17 | 0.39 | 0.56 | 0.80 | 5.00 | 0.60 |

**Table 2:** Summary statistics of the consideredvariables (* variables are in the logarithmic form. ° variables are divided by the population i.e., per capita).

The variables referring to the origin country are indicated with the subscript "$_O$", while the variables regarding the destination are reported with the "$_D$" subscript. The remaining variables are related to origin- destination combinations. Focusing on the environmental initiatives' variables, there are more than twice the amount of public environmental initiatives (with a mean of 0.22) than private initiatives (with a mean of 0.56) per province. Moreover, private initiatives are much more spatially clustered as indicated by the two medians.

Figure 1 shows the spatial distribution of tourism green initiatives over the Italian peninsula. Again, we keep the distinction between the initiatives undertaken by firms and policies applied by municipalities.

Again, it is remarkable that, although the green initiatives undertaken by the provinces are more distributed throughout the country, the green initiatives and labels to which tourism businesses have adhered are more concentrated in Northern Italy and around the capital, which correlates with higher economy activity.

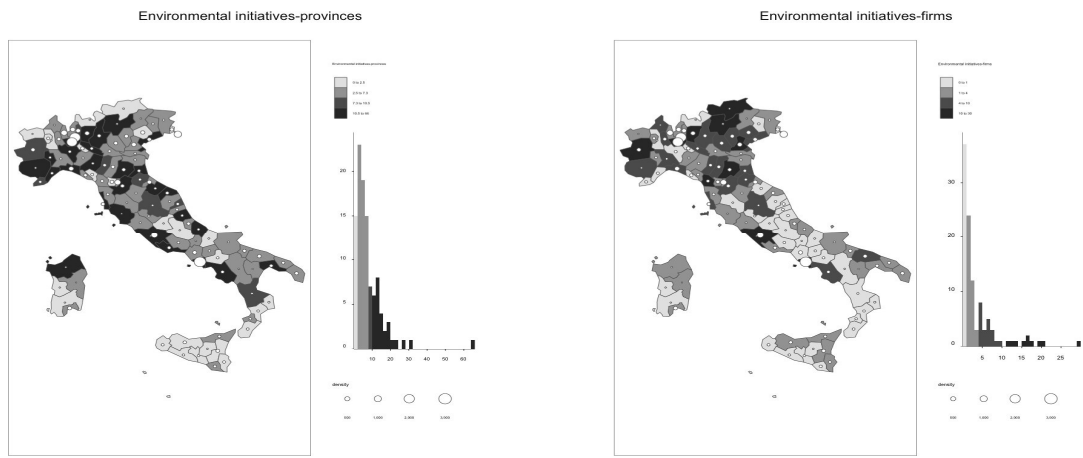

**Figure 1:** Spatial distribution of tourism green initiatives undertaken by the public authorities and private firms over the Italian peninsula compared to the population density.

# 5 Results

The results of the estimated gravity models are reported in Table 3.

*Gravity Models Results*

| | Dependent variable: Tourism flows | | | |
|---|---|---|---|---|
| | (1) | (2) | (3) | (4 - Fixed Effects) |
| Distance | −0.720*** (−0.820, −0.610) | −0.300*** (−0.400, −0.210) | −0.310*** (−0.400, −0.210) | −0.550*** (−0.690, −0.420) |
| GDP-origin | 0.410*** (0.300, 0.520) | 0.470*** (0.380, 0.560) | 0.470*** (0.380, 0.560) | 0.760*** (0.480, 1.000) |
| GDP-destination | 2.400*** (2.100, 2.700) | 1.800*** (1.300, 2.200) | 1.500*** (1.100, 2.000) | 1.500*** (1.200, 1.800) |
| Unemployment | | −4.200*** (−6.400, −2.100) | −4.200*** (−6.400, −2.000) | −4.000*** (−5.400, −2.600) |
| Density | | −0.210*** (−0.320, −0.098) | −0.190*** (−0.310, −0.080) | −0.200*** (−0.270, −0.120) |
| Transport accessibility-destination | | 0.620*** (0.290, 0.950) | 0.640*** (0.300, 0.970) | 0.630*** (0.420, 0.840) |
| Museum visitors | | 0.520*** (0.450, 0.590) | 0.500*** (0.420, 0.570) | 0.500*** (0.450, 0.550) |
| Coasts (km) | | 0.200*** (0.160, 0.230) | 0.200*** (0.170, 0.240) | 0.200*** (0.180, 0.230) |
| Domestic flows | | 5.000*** (4.600, 5.300) | 5.000*** (4.600, 5.300) | 4.400*** (4.100, 4.700) |
| N° tourism firms | | 0.720*** (0.630, 0.820) | 0.640*** (0.530, 0.740) | 0.640*** (0.570, 0.700) |
| Air pollution | | −0.930*** (−1.300, −0.600) | −0.990*** (−1.300, −0.660) | −0.990*** (−1.200, −0.790) |
| Environmental initiatives-firms | | | 0.360*** (0.120, 0.610) | 0.370*** (0.210, 0.530) |
| Environmental initiatives-province | | | 0.095 (−0.021, 0.210) | 0.094** (0.021, 0.170) |
| Constant | −18.000*** (−21.000, −15.000) | −19.000*** (−23.000, −14.000) | −15.000*** (−20.000, −11.000) | −17.000*** (−22.000, −13.000) |
| Observations | 3,424 | 3,424 | 3,424 | 3,424 |
| $R^2$ | 0.160 | 0.480 | 0.480 | 0.790 |
| Adjusted $R^2$ | 0.160 | 0.480 | 0.480 | 0.790 |
| Residual Std. Error | 2.200 (df = 3420) | 1.700 (df = 3412) | 1.700 (df = 3410) | 1.100 (df = 3381) |
| F Statistic | 221.000*** (df = 3; 3420) | 285.000*** (df = 11; 3412) | 243.000*** (df = 13; 3410) | 309.000*** (df = 42; 3381) |
| Note: | | | | *p<0.1; **p<0.05; ***p<0.01 |

**Table 3:** Number of provinces: 107; number of countries 32; total number of observations: 3,424. Estimation method: Ordinary Least Squares with White heteroskedasticity-consistent standard errors. Tourism flows, GDP's, density, transport accessibility, Museum visitors, coasts, n° of tourism firms and air pollution are log-transformed. In parentheses, 95% confidence intervals are reported.

The first column exhibits the results of a basic gravity model with masses (GDP per capita of origin and destination) and distance as a resistance term (for a robustness check, please see Table 4 in Appendix). The estimates confirm consolidated findings in the literature: as the distance between two places increases, the expected tourism flow decreases. On the contrary, the elasticity of GDP is positive and highly significant for both origin (0.41) and destination (2.4). However, the income elasticity in the origin is estimated to be lower than one (see Roselló Nadal and Santana Gallego, 2022), indicating that tourism is a normal good and not a luxury one. The high elasticity for the destination, instead, suggests that the positive association between tourism incoming flows and destination economy is strong (Marrocu and Paci, 2013). Considering the distance, the coefficient is significantly negative but higher than −1, meaning that the effect is less powerful than previous results. This could be due to the inclusion of domestic incoming flows in each province, taking as a centroid the geographical center of Italy, which is closer than other origins for most of the destinations.

In the second model, we added three variables and four controls, and a binary variable for domestic flows to deepen the pulling forces affecting tourists' choices. We found that unemployment, population density, and air pollution are negatively correlated with incoming tourism flows, suggesting that (*i*) people tend to visit less populated areas compared with heavily populated, (*ii*) unemployment is higher in Italian provinces less involved in tourism activities, and (*iii*) tourists, on average, tend to prefer less polluted cities (confirming previous findings in the literature, as, e.g., Xu and Dong, 2020). Also, the number of tourism firms located in the destination suggests that higher flows are positively associated with a higher firms' presence. However, the elasticity is lower than one, meaning that the relationship is less than proportional. Additionally, we included a measure of transport accessibility as a proxy for transport infrastructure, the number of visitors in museums, and the km of coasts as a proxy for the attractiveness of the destination. All these regressors exhibit positive coefficients and highly significant levels.

The third and fourth model introduces the main variables of this study, namely we include the environmental sustainability of destinations and the fixed effects for the country of origin. First, it is noteworthy that both the environmentally sustainable attitude of firms operating in the provinces and the initiatives and labels achieved by the local authorities in the provided services show a positive effect and that the coefficients are significantly different from zero.

Regarding the environmental facets affecting the provinces' institutions, its coefficient is not significant in the third model. However, once fixed effects for the origin countries are included, controlling for any permanent or persistent differences in the behavior of tourists over the considered European states, this variable becomes statistically significantly different from zero. Moreover, it is important to note how the estimates remain consistent over the last two models, with distance as the only exception, changing from −0.30 to −0.53. These results indicate a clear association between tourism demand and the environmental sustainability of the destination endowment, suggesting that sustainable facets should be part of the determinants of modern models in tourism demand theories, especially when a gravity specification is considered. Hence, although both private and public environmental measures and labels positively affect tourism incoming flows, an important consideration about their magnitude should not be overlooked. With respect to this, the magnitude of firms' environmental involvement is estimated to be around 35.1% higher than that of public policies[1]. The *marginal* effect of one additional private initiative—being 45%—is much stronger associated with tourism flows though than one additional public initiative—which is 9.9%. This means that although the commitment to sustainability in tourism is likely to be positively related to tourism flow, the importance of private firms' commitment seems to be stronger than that of public efforts.

The results from our preferred model (4 - Fixed Effects) point out to the following observations. First, tourism firms such as hotels and accommodation facilities are more likely to directly influence tourists' destination decisions than the municipalities themselves, thanks to their major investments in marketing activities. Strong commitment of the local public authority, focused on improving sustainability at provincial level, is not only limited to promoting tourism but also refers to the needs of the local population. Second, the communicative power of virtuous areas (in terms of sustainable investments and sustainable tourism infrastructure) can be confounded when combined with other destination regions. On the other hand, marketing activities of businesses are more likely to directly impact tourists and their decision-making. Third, and finally, our findings strongly suggest that both private and public environmental measures and labels have a positive effect on tourism incoming flows, although importance of private firms' commitment is stronger than the public efforts.

# 6 Limitations and future researches

The conducted analyses are based on cross-sectional data. This choice was driven by two reasons. First, after 2019 the tourism sector has experienced severe instability due to the Covid-19 pandemic which would add a confounding factor to the model. Second, it is only in recent years that environmental policies and labels have seen a large diffusion, so the phenomenon is scarce in historical data (time-series data for the considered environmental variables are not available).

Furthermore, as emphasised by several contributions in the literature (see, among others, Mycoo, 2006), the lack of effective and reliable tourism sustainability indicators severely limits the incisiveness of sustainability policies introduced by legislators and thus the reporting of data in a timely manner. This, coupled with disagreements in the very definition of sustainable policies and sustainable tourism policies (e.g., Guo et al., 2019; Farsari et al., 2007), led us to favour the deterministic detail and informational strength of the data on the sustainability measures in existence considered in this paper, at the expense of a focus on more complex models (such as panel data analyses). Opting for multi-year analyses offers the opportunity to address the possible presence of endogeneity, yet in order to answer the research question of this paper, adopting an empirical approach, would have required the use of proxy variables to capture the evolution of green labels concerning the Italian tourism sector, necessarily introducing noise and reducing the exploratory power of this research.

Moreover, the use of cross-sectional data may preclude causal inference and limits the ability to capture the temporal dynamics of the relationship between tourism development and sustainability.

---

[1] We calculate this at the means of both environmental initiatives' variables, being 0.22 and 0.56 for private and public initiatives, respectively.

Nevertheless, the inclusion of country specific origin fixed-effects partially controls for the presence of unobserved heterogeneity among the countries of origin, which could influence the flows of tourists towards a destination. This is important as there are many unobserved factors that might vary between different countries of origin, such as cultural differences or tourism preferences. Including fixed effects for the countries of origin allows to control for these systematic differences over the countries of origin and to focus on the effects of the variables of interest (i.e., environmental variables). Nonetheless, a use of spatially more disaggregated data may lead to an improvement in causal inference.

# 7 Conclusions

This study investigates the determinants of tourism incoming flows in Italy in 2019. The paper focuses on the pull forces, with particular attention to environmental aspects. By means of gravity models of increasing completeness, the present study constitutes a seminal attempt to examine the complex relationship between tourism demand and environmental sustainability.

The obtained results reveal a positive connection between sustainable initiatives and tourism incoming flows. Our findings indicate that the efforts of businesses have a greater impact on tourism demand than municipal labels, suggesting that eco-labels such as EMAS, ECOLABEL, BioHotels, and GreenKey are appreciated by tourists and positively influence their destination decision-making process. The models show a negative relationship between tourism flows, distance between origin and destination, unemployment, population density, and air pollution. In contrast, positive coefficients were found for GDP, transport accessibility, museums' attractiveness, length of coastal areas, and number of tourism firms operating in the destination.

The results highlight the relevance of considering environmental sustainability as an important factor for modeling tourism demand, particularly when using gravity models. Our findings provide valuable insights for tourism policymakers and businesses, suggesting that efforts to promote sustainability in the tourism sector could have significant benefits for both the industry and the wider economy. Specifically, private initiatives have greater impact compared to public initiatives, so that a policy effort focused on stimulating and facilitating private initiatives may be preferable.

## Declaration of interest

The authors declare that they have no known competing financial interests or personal relationships that could have appeared to influence the work reported in this article.

# Appendix

|  | Gravity Models Results PPML | | | |
|---|---|---|---|---|
|  | Dependent variable: | | | |
|  | Tourism flows | | | |
|  | (1) | (2) | (3) | (4 - Fixed Effects) |
| Distance | −1.400*** (−1.500, −1.300) | −0.310*** (−0.440, −0.180) | −0.360*** (−0.490, −0.230) | −0.260*** (−0.340, −0.180) |
| GDP-origin | −0.033 (−0.200, 0.140) | 0.560*** (0.380, 0.740) | 0.550*** (0.370, 0.720) | 1.200*** (0.650, 1.700) |
| GDP-destination | 2.000*** (1.600, 2.500) | 1.900*** (1.200, 2.600) | 1.300*** (0.570, 2.000) | 1.300*** (0.920, 1.600) |
| Unemployment |  | −4.100** (−7.900, −0.310) | −3.700* (−7.400, 0.056) | −4.200*** (−5.900, −2.400) |
| Density |  | −0.500*** (−0.680, −0.320) | −0.510*** (−0.690, −0.330) | −0.500*** (−0.590, −0.420) |
| Transport accessibility-destination |  | −0.340 (−0.900, 0.220) | −0.076 (−0.630, 0.480) | −0.120 (−0.380, 0.140) |
| Museum visitors |  | 0.350*** (0.230, 0.470) | 0.340*** (0.220, 0.470) | 0.340*** (0.290, 0.400) |
| Coasts (km) |  | 0.180*** (0.130, 0.230) | 0.190*** (0.140, 0.240) | 0.190*** (0.170, 0.220) |
| Domestic flows |  | 3.200*** (3.000, 3.500) | 3.200*** (2.900, 3.400) | 3.400*** (3.200, 3.700) |
| N° tourism firms |  | 0.400*** (0.260, 0.550) | 0.240*** (0.082, 0.400) | 0.270*** (0.190, 0.340) |
| Air pollution |  | 0.460* (−0.042, 0.960) | 0.340 (−0.160, 0.840) | 0.330*** (0.094, 0.560) |
| Environmental initiatives-firms |  |  | 0.540*** (0.250, 0.820) | 0.510*** (0.380, 0.640) |
| Environmental initiatives-province |  |  | 0.095* (−0.007, 0.200) | 0.091*** (0.044, 0.140) |
| Constant | −6.000** (−11.000, −1.400) | −15.000*** (−22.000, −6.700) | −8.400** (−16.000, −0.350) | −15.000*** (−21.000, −8.100) |
| Observations | 3,424 | 3,424 | 3,424 | 3,424 |
| Pseudo $R^2$ | 0.37 | 0.7 | 0.7 | 0.84 |

Note: *p<0.1; **p<0.05; ***p<0.01

**Table 4:** Number of provinces: 107; number of countries 32; total number of observations: 3,424. Estimation method: Poisson-pseudo MLE (loglink) with White heteroskedasticity-consistent standard errors. Tourism flows, GDP's, density, transport accessibility, Museum visitors, coasts, n° of tourism firms and air pollution are log-transformed. In parentheses, 95% confidence intervals are reported.